# The Online Behaviour of the Algerian Abusers in Social Media Networks

Kheireddine Abainia

PIMIS Laboratory, Department of Electronics and Telecommunications, Université 8 Mai 1945 Guelma, 24000, Guelma, Algeria
`k.abainia@univ-guelma.dz`

**Abstract.** Connecting to social media networks becomes a daily task for the majority of people around the world, and the amount of shared information is growing exponentially. Thus, controlling the way in which people communicate is necessary, in order to protect them from disorientation, conflicts, aggressions, etc. In this paper, we conduct a statistical study on the cyber-bullying and the abusive content in social media (i.e. Facebook), where we try to spot the online behaviour of the abusers in the Algerian community. More specifically, we have involved 200 Facebook users from different regions among 600 to carry out this study. The aim of this investigation is to aid automatic systems of abuse detection to take decision by incorporating the online activity. Abuse detection systems require a large amount of data to perform better on such kind of texts (i.e. unstructured and informal texts), and this is due to the lack of standard orthography, where there are various Algerian dialects and languages spoken.

**Keywords:** Offensive language, Abusive content, Cyber-bullying, Algerian dialectal Arabic, Social media.

## 1 Introduction

The emergence of social media networks has substantially changed people's life, where this kind of websites became the parallel life of some people around the world. In Algeria, Facebook social media is the leading and the most visited social media website, in which people share news and moments, and express ideas as well. Accordingly, it becomes one of the main sources of information, news, education and culture (Abainia, 2020). As an outcome, some ill-intentioned people take advantage to broadcast misinformation for different malicious intents (e.g. threatening governmental officers, celebrities, etc.).

Despite social media networks have a positive influence on people's life, they might have a negative outcome like the cyber-criminality. Indeed, it was reported by Dailymail[1] that the British police receives one Facebook cyber-crime each 40 minutes, and 12 hundreds cyber-criminalities have been logged in 2011 (Abainia, 2020). Several cyber-crimes may lead to suicide or killing such as frauds, cyber-bullying, stalking, robbery, identity theft, defamation and harassment.

---

[1] http://www.dailymail.co.uk/

In this investigation, we conduct a statistical study to understand the behaviour of the Algerian online abusers (i.e. cyber-bullyers). The purpose of this study is to spot the abuser accounts from their activities and their interactions, in order to help automatic systems of abuse detection in making decisions. Indeed, the abusive content may take different forms such as a hate speech, harassment, sexisme, profanity, etc. Golbeck et al. (2017) defined five categories of harassment when they created a large corpus, namely: the very worst, threats, hate speech, direct harassment and potentially offensive. In this work, we only address the profanity (unreadable comments), while the harassment is not the purpose of this work.

We particularly focus on the Algerian online community, because Algeria is rich of linguistic varieties in one hand, i.e. several Arabic dialects and languages. On the other hand, the dialectal Arabic lacks of standard orthography, and could be written in Arabic script or Latin script (Arabizi). Generally, social media users simulate the phonetic pronunciation of the words in their writing. For instance, we may find consecutive repetitions of the same letter depending on some emotional passages, like the use of "*bzzzzzzzzf* " instead of "*bzf*" (meaning "*a lot*").

Beside the lack of standard orthography, we find the code-switching phenomenon, which consists of mixing more languages in the same sentence or conversation (Joshi, 1982). Usually, it occurs by multi-lingual speakers depending on the situation and the topic (Kachru, 1977). Hence, in the Algerian community, we spot different scenarios of code-switching such as Arabic-French, Arabic-English, Arabic-French-English, Arabic-Berber, Berber-French, Berber-English, Berber-French-English, and Arabic-Berber-French-English (very rare).

The most of works on abusive content detection are using machine learning or deep learning tools trained on a set of data, or using linguistic features like a lexicon of abusive words. However, these methods may have drawbacks in the case of the Algerian dialectal Arabic, because this task requires a huge training data covering different writing possibilities and a large lexicon with different patterns of abusive words and sentences. In order to overcome these issues, we have conducted a statistical study to spot the behaviour of the Algerian online abusers. Thus, we could predict potential user profiles that post abusive comments in social media networks. The ground truth of this study is based on several Facebook user profiles, where 200 differents users from different regions have been involved to study their activities among 600 Algerian abusers.

## 2    Related work

In this section, we highlight some research works carried out on the abusive and harassing content, where we state related works from the sociological viewpoint and the computational linguistics as well.

Al Omoush et al. (2012) studied the impact of Arab cultural values on Facebook, where they investigated to study the motivation, attitudes, usage patterns and the continuity of membership values in social networks. The study revealed that social media networks break down restriction barriers in front of Arabic young people, where the latter face a lot of cultural, social, religious, moral and political restrictions (Al Omoush et al., 2012).

Awan studied the online hate speech against the Islamic community in the United Kingdom (Awan, 2014). The author analysed 500 tweets written by different users, i.e. 100 users, where the statistics showed that 72% of the tweets were posted by males. The author categorized offensive users into several categories, and among these categories the reactive and accessory reported high statistics (95 and 82 cases, respectively). Reactive users are persons following major incidents and take the opportunity, while the accessory category represents persons joining a hate conversation to target vulnerable people (Awan, 2014). Later, the author studied the same phenomenon on Facebook, where the author analysed 100 different Facebook pages (Awan, 2016). The analysis figured out that 494 comments contain hate speech against the Muslim's community in the UK, where 80% of the comments were written by males. In addition, the author categorized the offenders into five different categories, i.e. opportunistic, deceptive, fantasist, producer and distributor. From the two studies, in overall the offenders used the same keywords and hashtags such as: Muzrats, Paedo, Paki, ISIS, etc. (Awan, 2014; Awan, 2016).

Lowry et al. (2016) studied the adults' cyber-bullying and the reasons for cyber-bullying, while the most of research works were focused on the adolescents. The authors particularly proposed an improved social media cyber-bullying model, where they incorporated the anonymity concept with several features in the learning process of cyber-bullying.

The abusers content in community question answering was studied in (Kayes et al., 2015), where the authors focused on the flagship of inappropriate content by analysing 10 million flags related to 1.5 million users. The analysis figured out that the most of the flags were correct, and the deviant users receiving more flags get more replies than ordinary users.

Arafa and Senosy studied the cyber-bullying patterns on 6,740 Egyptian university students in Beni-Suef (Erafa and Senosy, 2017). The questionnaire responses showed that 79.8% of females receive a harassment, while 51.8% of males receive a flaming content. Moreover, among the victim feelings, the anger, hatred and sorrow are common to the most of the victims. The data analysis figured out that the students from rural areas and medicine students are less exposed to cyber-bullying (polite and respect moral values). Conversely, the students living in urban areas and sociology students are more exposed to cyber-bullying. A similar study has been conducted on a group of Saudian students, where 287 students were selected as a case study among 300 responding to a questionnaire (Al-Zahrani, 2015).

The most of cyber-bullying studies addressed adolescents and teenagers. For instance, a group of Arab teenagers (114 teenagers) living in Israel was selected to study the cyber-bullying and the relationships with emotional aspects (Heiman and Olenik-Shemesh, 2016). The study reported that 80% of the students are subject to different forms of cyber-bullying such as: spreading offensive rumours, harassment, humiliation regarding the physical appearance, and sending sexual content. In addition, the study also figured out that the cyber-bullying victimization conducts to loneliness and anxiety. Another study comparing the cyber-bullying of Jewish and Arab adolescents in Israel was conducted in (Lapidot-Lefler and Hosri, 2016). The analysis showed that Jewish students use the Internet more frequently than Arab students, and consequently the latters are less exposed to cyber-criminality. Moreover,

Arabic females are more bullied than males in contrast to Jewish students (i.e. no difference between genders).

In contrast to the above studies, Triantoro studied the cyber-bullying phenomenon on a group of 150 high school students in Jogjakarta (Triantoro, 2015). The Triantoro's study showed that 60% of the students are not victims of cyber-bullying.

Tahamtan and Huang conducted a statistical study on the cyber-bullying prevention, for which they collected over 6k tweets (Tahamtan and Huang, 2019). The authors concluded that parents and teachers should be trained to know different aspects of cyber-bullying, and how to prevent this act in schools.

Finally, Almenayes studied the relationship between cyber-bullying and depression, and he particularly focused on the gender and the age (Almenayes, 2017). The statistical study showed that females are more likely to have depression in contrast to males, while the age is not a good predictor of depression.

## 3    Algerian Social Media Users

The emergence of social media networks in Algeria considerably increases the code-switching phenomenon, where it also occurs by monolingual and illiterate people, because the use of foreign languages (e.g. French) is seen as a prestigious and elegant way to communicate with people (Abainia, 2020). Thus, monolingual people can acquire some French words from the spoken language and employ them within a sentence like "frr tjr bien et en **bon sen**té" (correct sentence is "frr tjr bien et en **bonne santé**").

```
Avantage IMTIYAZ 12Go INTERNET+HADRA wa SMS ILLIMITES li
Djezzy+1300DA li echabakat el wataniya el okhra b'1000DA/30
      youm! 3ard salih hata 08/04. Tape*444*15#
```

**Fig. 1.** Offer promotion by an Algerian ATM company. The bold words are French words

Regarding the code-switching expansion, some ATM companies use Arabic-French code-switching to promote offers (Figure 1).

### 3.1    Online social behaviour

In our previous work (Abainia, 2020), we have noticed that several Algerian users often use pseudonyms reflecting their personalities or imaginations. For instance, if a user feels resilient and stronger, he may choose the pseudonym "*Gladiator*" reflecting a strong personality. In addition, some users write their surnames or pseudonyms using accented characters (e.g. Õ, Ŭ, ʾA, Ơ, ʾH, etc.) not belonging to the French character set that contains 26 letters and some vowels (é, è, à, ï and î).

Several social media users spread fake information about them for various reasons (e.g. keeping anonymity). For instance, some Algerian users make European countries as their living country, because they want to migrate there and they always publish and share status and images about this country. Furthermore, some Algerian geeks make renowned universities and institutions (e.g. MIT) as their studying institutions (Abainia, 2020).

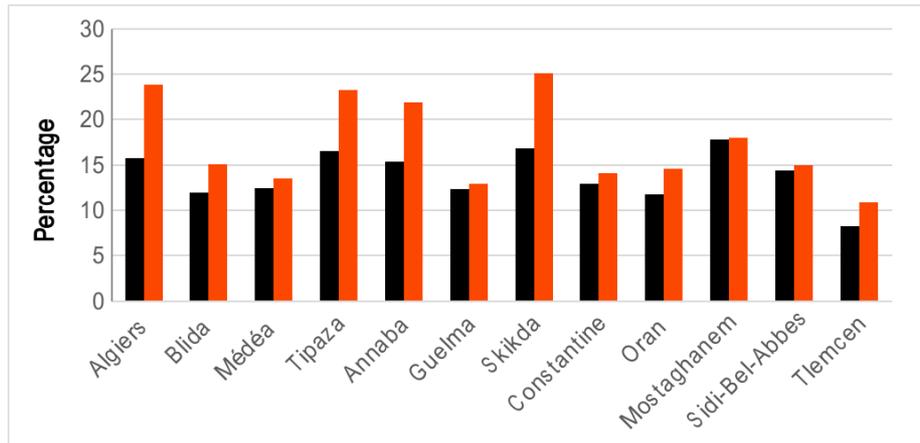

**Fig. 2.** Percentage of French words used by both genders through different cities. Blue peaks are male percentages and red peaks are female percentage.

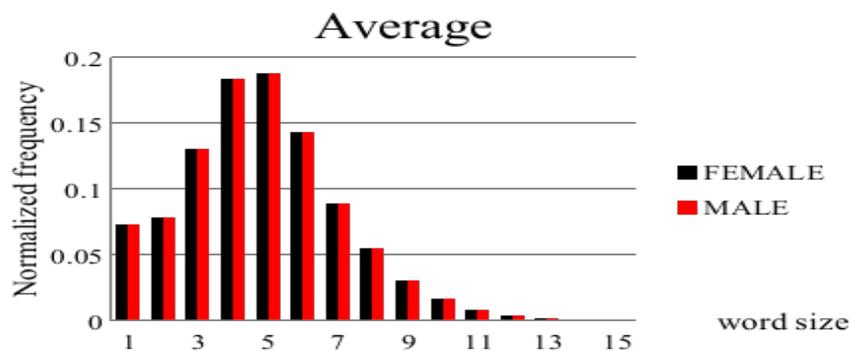

**Fig. 3.** Distribution of word lengths over 2400 Facebook comments in DZDC12 corpus

Finally, we assume that there are two kinds of social media users: naive and rational. Naive users are generally illiterate and little minded, as well as they have the same writing style, the same spoken language and they believe everything. Conversely, rational people change the writing style depending on the context and do not believe everything, as well as they are more polite and less aggressive.

### 3.2 Writing behaviour

The Algerian community code-switch more frequently in social media, especially between Arabic-French, because French is the second language. From our previous analysis undergone on DZDC12 corpus (i.e. 2.4k texts written by males and females), females code-switch more frequently than males over all the cities (i.e. 12 cities), and especially in Algiers, Blida, Tipaza, Annaba, Skikda and Oran, where the difference is clear enough (Abainia, 2020). In addition, it was noticed that misspelling errors commonly occur by males, while the latter less use abbreviations.

Algerian social media users write in Arabic script and Arabizi[2] interchangeably, and the latter has irregular orthography. However, in southern cities the community commonly uses the Arabic script.

Figure 3 aims to report some statistics about the word lengths used by both genders. Surprisingly, the two genders produce the same distribution, where the peaks drop to the ground after 13 letters. In addition, four-literal and five-literal words produce the highest peak, and uni-literal words are also used in Algerian Arabizi (Abainia, 2020). For the latter case, some bi-literal and tri-literal words are often abbreviated to a single letter (e.g. *"fi"* or *"fe"* meaning *"in"* are abbreviated to *"f"*).

## 4    Algerian online abusers behaviour

To conduct this study, we have manually crawled 600 Facebook abusive comments written by different users (i.e. 600 Facebook users), where we have ignored the comments written by the same users. The comments were collected from public pages and groups related to different topics such as news, anti-system and politics, advertise, humour, music & rap, volunteering, stories and diverse. It was noticed that among the 600 comments, males, excepting one female, wrote all the comments.

From the collected data, we have randomly selected 200 users to analyze their profiles and gathering their information. We have saved the abuser's account name (and the profile's link), post text, illustrative media (photo or video), page category, post subject (i.e. politics, news, joking, etc.), post abusiveness (yes or no), post with hashtag (yes or no), abusive comment, abuser agreement with the post (support or against), number of reactions to the abusive comment. In addition, the abuser's account name with special characters (yes or no), abuser's gender (male or female), abuser's account activities (publication categories), average number of reactions to the abuser publications, average number of comments to the abuser publications, visited local regions (yes or no), and finally the visited countries (yes or no).

### 4.1    Statistics

Among the involved users (i.e. 200), 43% of them use their real names and the remaining use pseudonyms. For the latters, we have noticed surnames combined with nicknames (e.g. *"Mohamed Lakikza"*, *"Samir Babelwad"*, etc.), and pseudonyms reflecting celebrities and imaginations (e.g. *"Prince Charmant"*). Moreover, among the 43% users with real names, three of them use special characters, whereas 29 users use special characters among the other 57% (pseudonyms).

Eighteen users among 32 using special characters (both categories) share selfies, 11 share sports images and news and 11 share status and images of heartbroken (Figure 4).

---

[2] Arabic words written in Latin script

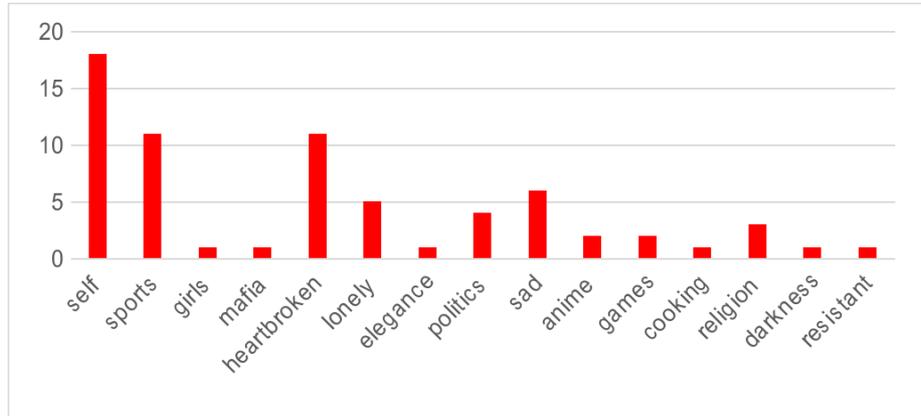

**Fig. 4.** Statistics about the abusers using special characters

As depicted in Figure 5, the most of abusive comments are found in anti-system and diverse pages (94 and 71 comments, respectively). In addition, the subject of the second category (i.e. diverse) is joking or politics, where the first subject may result a sarcasme and abusiveness. Nevertheless, since the end of 2018, the Algerian political situation is instable and the most of Algerian people protest against the governmental staff in streets and social media. The fact of sharing political subjects involves hate and angry, which may provoke to post abusive content.

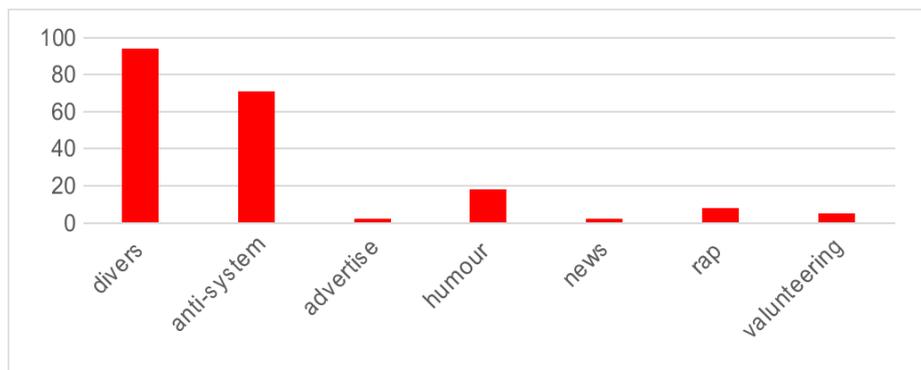

**Fig. 5.** Number of comments pert page category

On the other hand, the most of Algerian community are religiously preservative, and the fact of seeing controversial things to their beliefs involves angry and hate. For example, a video showing some Algerian women dancing in the street involves a hateful reaction, because in the Algerian beliefs, women should not be exposed without veil and should not do immoral acts like dancing outside.

By analyzing the abuser profiles and their activities, in overall, they do not receive much feedback and reactions from their connected peers, excepting a few ones. In particular, 24.5% of the abusers receive more reactions (between 70 and 300), while the majority of the others receive between 0-10 of reactions for their posts. Among the 24.5% of the abusers, some ones receive until 300 comments (they share selfies alone or with friends), and the majority receives between 10-20 comments. In contrast to the users not receiving many reactions (75.5%), they also did not receive comments (0 comments in almost the cases).

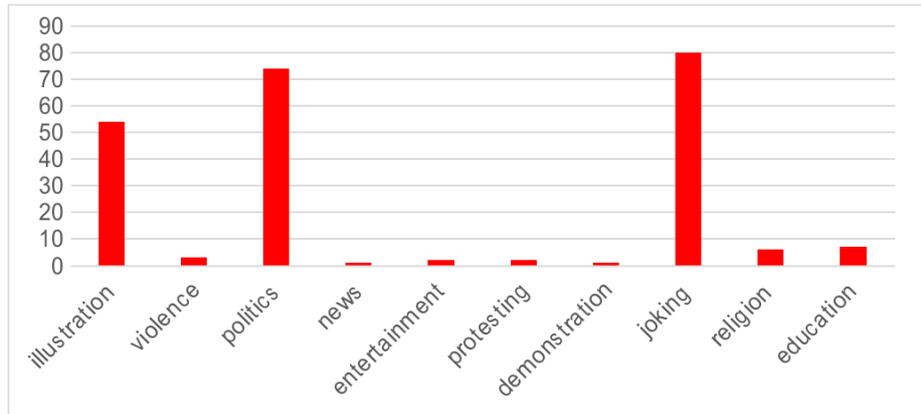

**Fig. 6.** Number of abusive comments per post category

Figure 6 depicts the number of abusive comments per each post category, where the comments may belong to two subjects. From Figure 6, we notice that political and joking subjects involve abusiveness, because as stated above the most of Algerian people are against the current government. Moreover, we also notice that educational and religious subjects involve abusiveness.

Algerian people are religiously preservative and restrictive to the religion (i.e. *Islam*) and education principles, and if their religion is insulted it would make them angry and lead to abusiveness. In addition, insulting the moral principles and the education quality may provoke negative reactions (abusiveness). Finally, it is obvious that joking subjects (e.g. sarcasm) raises abusiveness.

## 4.2 Abusers activities on their accounts

Figure 7 depicts the number of abusers per activity, where the same abuser may do multiple activities (e.g. selfie, sadness, etc.). From the figure, the most of the abusers share selfies (alone or with friends and family) and Football clubs and athletes. Starcevic et al. (2018) have studied the medical terminology of the selfie addiction in social media, where the authors have defined the phenomenon as a mental disorder, which consists of an obsessive need to post selfies in social media, and requires a professional treatment. We assume that they may have a mental illness and try to draw attention of their peers. De Choudhury et al. (2013) addressed the depression prediction in social media, where the authors used a clinical survey model via Amazon Mechanical Turk. Among the negative emotions related to depression, the depressed users share feelings of worthlessness, guilt, helplessness, and self-hatred.

On the other hand, we have noticed another form of sadness, which consists of sharing religious posts, because conservative Islamic people cannot post profanity and abusiveness (i.e. prohibited in their beliefs). In overall, such users do not share joking, sports and political posts. Conversely, the abusers in our dataset share religious posts combined with political, joking, sports, sadly and violence posts. This kind of abusers has used a direct harassment to abuse others, where we classify these comments as "*the very worst*" according to Golbeck et al. (2017).

It is worth mentioning that sport posts are generally concerning Football (i.e. national and international clubs), because the latter is considered as favorite sport of the Algerian community (and the Arabic world in overall). More frequently, the users share selfies in the stadiums and some pictures of their preferable players and clubs (i.e. *Real Madrid*, *Barcelona*, etc.). Usually, the Algerian football supporters (especially who go to the stadium) learn abusive words and abusive language, because they usually sing abusively in the stadium.

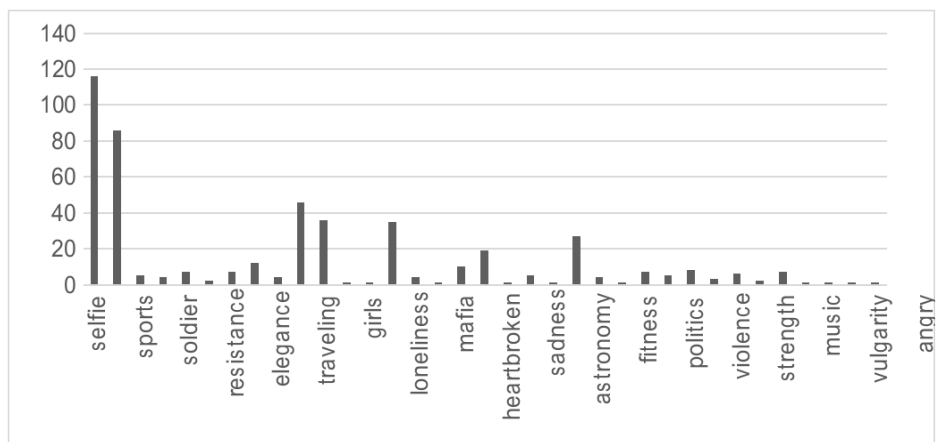

**Fig. 7.** Number of abusers per activity

We have noticed that almost the users sharing sports posts receive more reactions, i.e. up to 20 comments and 200 likes. Moreover, the users sharing travelling posts also receive more reactions, i.e. until 15 comments and 100 likes. Nevertheless, for the other activities, the number of reactions varies from a user to another. Thus, we could not draw a general conclusion, because the number of reactions depends to the user tie and engagement.

### 4.3   Writing style

In overall, the language script (Arabic or Latin) of the abusive comments is independent to the script of the admin's post. However, we have noticed that the most of abusers write in Arabic script to reply to political subjects. Moreover, some users whose names are written in Arabic script always write in the same script. Conversely, we cannot predict the language script based on other information like the pseudo/name and the use of special characters in the names.

Among the 200 involved abusers, 78.5% of them are against the admin's posts, 15% agree and the remaining are neutral. The most of the comments against the posts are directly targeting the admin when the latter is commenting something (especially a political situation). On the other hand, they are targeting the subject (i.e. person stated in the post content) in the case of sharing news. From figure 8, the most of the abusers write short comments (below 8 words), as well as some abusers write two or three words (i.e. abusive words).

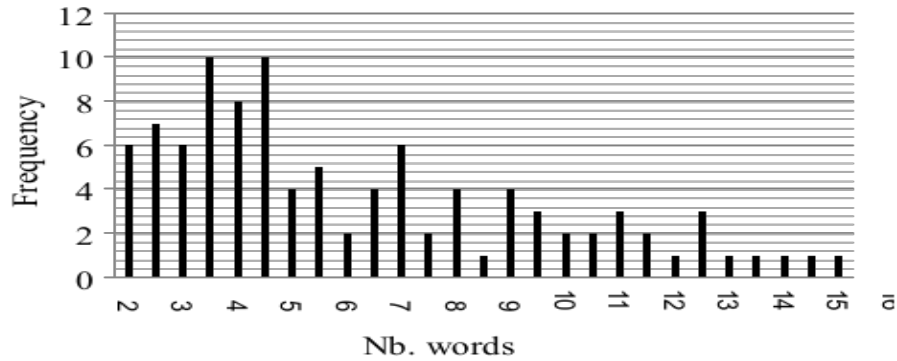

**Fig. 8.** Number of abusers per user's activity

Table 1. Examples of some indecent words retrieved from our dataset

| N° | Arabic word | Buckwalter Tansliteration |
|---|---|---|
| 1 | الكرفة | Alkrfp |
| 2 | يضربولو ترمه | yDrbwlw trmh |
| 3 | Nike moke | - |
| 4 | hetchoun yemak | - |
| 5 | hatchoun yamk | - |
| 6 | nik moook | - |
| 7 | zabo | - |
| 8 | yeniko | - |
| 9 | 7atchoun yamakk | - |
| 10 | nikamok | - |
| 11 | 9ahbon | - |
| 12 | أزبي | Azby |
| 13 | نتمنيكو | ntmnykw |
| 14 | نيكو ماتكم | nykw mAtkm |
| 15 | ولاد لقحاب | wlAd lqHAb |

Due to the irregular orthography of Arabizi and dialectal Arabic, it is obvious that each word has various writing varieties. However, we can regroup the words into three words, i.e. *nik* (meaning "*fuck*"), *zebi* (meaning "*penis*") and *ya3ti* (meaning "*prostitute*"). In addition, some users sometimes obfuscate indecent words by replacing some characters with others. For instance, the word "*zebi*" (زبي) becomes "*shepi*" (شبي) where the letter "*z*" is obfuscated with "*sh*" (pronounced like *shop*). Another example, the word "*tnaket*" (تناكت) is sometimes obfuscated as "*tnakhet*" (تناخت).

It is worth mentioning that sometimes the abusers could use classical Arabic to write an abuse, and in such case, the latter can be detected from the general context. For example, the second entry in Table 1 could be written as "يضربولو المؤخرة", where the word "ترمة" is indecent and the word "المؤخرة" is familiar and more polite, but the sentence remains abusive.

We notice from Table 1 that the 4th, 5th and 9th entries are the same sentence written differently, as well as we notice the same thing with the 3rd, 6th, 10th and 14th entries.

Finally, from the collected data, it is noticed that the abusers use the same writing style and the same words when they interact with others. For instance, when a friend share a selfie the abuser generally write the same comments "*rak 4444 hbibi*" (meaning "*you are the best buddy*"), "*tjrs 4 frr*" (meaning "*my brother, you are always the best*"), "*dima zine hbibi*" (meaning "*buddy, you are always handsome*"), etc.

## 5 Conclusion

In this investigation, we have addressed the problem of online cyber-bullying by focusing on the Algerian online abusers. In particular, we have highlighted the social behaviour of the Algerian online community, as well as the abusers' writing style and online activities.

The ground truth of this study is based on 200 abusive comments (contain profanity) written by different users (i.e. 200 users) from different Algerian regions. We have manually collected 600 comments in total written by different Algerian Facebook users, where the comments written by the same users were ignored. Among the 600 comments, we have arbitrarily selected 200 comments to inspect the user profiles and gathering the related information. We found that males wrote all the abusive comments, excepting one comment written by a female.

The statistics showed that 57% of the abusers use pseudonyms instead of real names, and 16% use special characters not belonging to the French character set. It was also noticed that political subjects in the most cases involve cyber-bullying and profanity, because the Algerian community is against the current governmental staff. In addition, it was noticed that the most of abusers are addicted with publishing selfies and sharing posts about sports, politics and negative feelings (e.g. sadness, heartbroken, etc.). Thus, we see from the collected data that the most of abusers are depressive, and we should be aware from this kind of users.

Because of the irregular orthography of Arabizi and dialectal Arabic, building a lexicon (of abusive words) covering different possibilities is tricky and requires more investigations. As perspective for future work, we would like to explore the writing variation across different Algerian regions and building a lexicon for abusive content.

## References

Abainia, K. (2020). DZDC12: a new multipurpose parallel Algerian Arabizi-French code-switched corpus, Journal of Langage Resources & Evaluation. VoL. 54, 2020, (pp. 419-455).

Almenayes, J. (2017). The Relationship between Cyberbullying Victimization and Depression: The Moderating Effects of Gender and Age. Journal of Social Networking, Vol. 6, No. 3, 2017, (pp. 215-223).

Al Omoush, K. S., Yaseen, S. G., and Alma'Aitah, M. A. (2012). The impact of Arab cultural values on online social networking: The case of Facebook. Journal of Computers in Human Behavior, 28(6), (pp. 2387-2399).

Al-Zahrani, A. M. (2015). Cyberbullying among Saudi's Higher-Education Students: Implications for Educators and Policymakers. World Journal of Education, Vol. 5, No. 3, 2015, (pp. 15-26).

Arafa, A., and Senosy, S. (2017). Pattern and correlates of cyberbullying victimization among Egyptian university students in Beni-Suef, Egypt. Journal of Egyptian Public Health Association, Vol. 92, No. 2, 2017, (pp. 107-115).

Awan, I. (2014). Islamophobia and Twitter: A Typology of Online Hate Against Muslims on Social Media. Policy & Internet, Vol. 6, No. 2, 2014, (pp. 133-150).

Awan, I. (June, 2016). Islamophobia on Social Media: A Qualitative Analysis of the Facebook's Walls of Hate. International Journal of Cyber Criminology, Vol. 10, No. 01, June, 2016, (pp. 1-20).

De Choudhury, M., Gamon, M., Counts, S., and Horvitz, E. (2013, June). Predicting depression via social media. Facebook: A Hidden Threat to Users' Life Satisfaction?. Proceedings of the 7th international AAAI conference on weblogs and social media.

Golbeck, J., Ashktorab, Z., Banjo, R. O., Berlinger, A., Bhagwan, S., Buntain, C., Cheakalos, P., A Geller, A., Gergory, Q., Gnanasekaran, R. K., Gunasekaran, R. R., Hoffman, K. M., Hottle, J., Jienjitlert, V., Khare, S., Lau, R., Martindale, M. J., Naik, S., Heather, L., Nixon, Ramachandran, P., Rogers, K. M., Rogers, L., Sarin, M. S., Shahane, G., Thanki, J., Vengataraman, P., Wan, Z., and Wu, D. M. (2017, June). A large labeled corpus for online harassment research. Proceedings of the 2017 ACM on Web Science Conference, (pp. 229-233).

Heiman, T., and Olenik-Shemesh, D. (2016). Computer-based communication and cyberbullying involvement in the sample of Arab teenagers. Education and Information Technologies, Vol. 21, No. 5, 2016, (pp. 1183-1196).

Joshi, A. K. (1982, July 05-10). Processing of sentences with intra-sentential code-switching. Proceedings of the 9th conference on Computational linguistics, (pp. 145-150), Prague, Czechoslovakia.

Kachru, B. B. (1977). Code-switching as a communicative strategy in India. In: M. Saville-Troike (ed.), Linguistics and anthropology. Georgetown University Round Table on Languages and Linguistics. Washington D.C.: Georgetown Univ. Press.

Kayes, I., Kourtellis, N., Quercia, D., Iamnitchi, A., and Bonchi, F. (May, 2015). The social world of content abusers in community question answering. Proceedings of the 24th International Conference on World Wide Web, International World Wide Web Conferences Steering Committee. May, 2015, (pp. 570-580).

Krasnova, H., Wenninger, H., Widjaja, T., and Buxmann, P. (March, 2013). Envy on Facebook: A Hidden Threat to Users' Life Satisfaction?. Proceedings of 11th International Conference on Wirtschaftsinformatik, February 27-March 1, 2013, Leipzig, Germany, (pp. 1-16).

Lapidot-Lefler, N., and Hosri, H. (2016). Cyberbullying in a diverse society: comparing Jewish and Arab adolescents in Israel through the lenses of individualistic versus collectivist cultures. Journal of Social Psychology of Education, Vol. 19, No. 3, 2016, (pp. 569-585).

Lowry, P. B., Zhang, J., Wang, C., and Siponen, M. (2016). Why do adults engage in cyberbullying on social media? An integration of online disinhibition and deindividuation effects with the social structure and social learning model. Information Systems Research, Vol. 27, No. 4, 2016, (pp. 962-986).

Starcevic, V., Billieux, J., and Schimmenti, A. (2018). Selfitis, selfie addiction, twitteritis: irresistible appeal of medical terminology for problematic behaviours in the digital age. Journal of Australian and New Zealand Psychiatry, Vol. 52, No. 5, (pp. 408-409).